\documentclass[aps,pra,twocolumn,groupedaddress,showpacs,showkeys]{revtex4}

\usepackage{graphicx}
\begin{document}
\vspace{1cm}
\newcommand{\be}{\begin{equation}}
\newcommand{\ee}{\end{equation}}

\title{Towards measuring variations of Casimir energy  by a superconducting cavity}

\author{ Giuseppe Bimonte, Enrico Calloni,  Giampiero Esposito,  Leopoldo Milano and Luigi Rosa}
\affiliation{Dipartimento di Scienze Fisiche, Universit\`{a} di
Napoli Federico II,   Via Cintia I-80126 Napoli, Italy; INFN,
Sezione di Napoli, Napoli, ITALY }

\date{\today}

\begin{abstract}
We consider a Casimir cavity, one  plate of which is a thin
superconducting film. We show that when the cavity is cooled below
the critical temperature for the onset of superconductivity, the
sharp variation (in the far infrared) of the reflection
coefficient of the film engenders a variation in the value of the
Casimir energy. Even though the relative variation in the Casimir
energy is very small, its  magnitude can be comparable to the
condensation energy of the superconducting film, and this gives
rise to a number of testable effects, including a significant
increase in the value of the critical magnetic field, required to
destroy the superconductivity of the film. The theoretical ground
is therefore prepared for the first experiment ever aimed at
measuring variations of the Casimir energy itself.

\end{abstract}


\maketitle

In recent years, new and exciting advances in experimental
techniques   have prompted a great revival of interest in the
Casimir effect. As is well known, this phenomenon is a
manifestation of the quantum zero-point fluctuations of the
electromagnetic field (for recent reviews, see \cite{bordag}). For
the simple case of two plane parallel, perfectly  conducting
plates of area $A$, separated by a distance $L$, the Casimir
energy is negative and equal to \be E^{(C)}=-\frac{
\pi^2}{720}\,\frac{\hbar c A}{ L^3}\;,\label{epara}\ee which
corresponds to an attractive force between the plates.

All experiments on the Casimir effect performed so far, measured
the Casimir force, in a number of different geometric
configurations. In this  Letter we find that by realizing a rigid
cavity, including a superconducting plate, it might be possible
for the first time to {\it measure directly variations of the
Casimir energy}. The basic idea is very simple: since the Casimir
energy depends on the reflective power of  plates,  there should
be a variation in the Casimir energy (and force), as soon as the
plate becomes superconducting,  because the transition determines
a sharp change in the reflective properties in the infrared (IR)
region. Indeed, an attempt at modulating the Casimir force by
changing the reflective power of  mirrors has been made recently
\cite{iann}, with negative results. This may appear as very
discouraging, especially if one considers that in this experiment,
based on the technology of hydrogen switchable mirrors,   there
was a large modulation in the optical region of the spectrum,
which is very relevant for typical submicron separations between
the mirrors. The possibility of success with superconducting
mirrors would seem even worse then,  since the superconducting
transition affects the reflective power only in the far IR region
\cite{glover}, which is clearly of little relevance for typical
Casimir cavities. There is however a very important difference
between our modulation scheme and the previous ones, which should
make it possible to obtain very high sensitivities. Indeed, we do
not mean to directly measure the relative variation of the Casimir
energy (or force) accompanying the transition, which we indeed
evaluate to be very small (typically, a few parts in hundred
millions or less, in our conditions). In the experimental setting
that we propose, aiming at a measurement of the critical field of
a thin superconducting film  included in a Casimir cavity, {\it
one is sensitive to variations of the Casimir energy as compared
with the condensation energy of the film}. This implies an
enormous improvement in sensitivity, for the condensation energy
is several orders of magnitude smaller than the Casimir energy,
such that even a tiny fractional variation in the latter can
produce significant effects on the critical field (see below).  We
observe another advantage of our setting, as the use of rigid
cavities allows  a large number of geometries, which will prove
useful in the study of the dependence of the Casimir effect on
geometry, indeed a distinctive feature of the Casimir effect,
arising from the long-range character of  retarded van der Waals
forces.

To be definite, we consider a double  cavity, consisting of two
identical plane parallel mirrors, made of a non-superconducting
and non-magnetic metal, between which    a plane superconducting
film of thickness $D= 5$ nm is placed, separated by an empty gap
of equal width $L=10$ nm from  the two mirrors.

For any  fixed temperature $T$ below the critical temperature
$T_c$, we wish to determine the shift in the value of  critical
parallel field $H_{c \|}(T)$ of the film, determined by the
Casimir energy of the cavity. Now, as  is well known, $H_{c
\|}(T)$ is determined by the difference between free energies
$\Delta F=F_{n}(T)-F_{s}(T)$ of the system (for zero field), in
the normal $(n)$ and in the superconducting $(s)$ state. For a
thin film ($D \ll \lambda, \xi$ with $\lambda$ the penetration
depth and $\xi$ the correlation length), exploiting known formulae
\cite{tink} we arrive at \be \left(\frac{H_{c
\|}(T)}{\rho}\right)^2 \,\frac{V}{8 \pi}=\Delta
F(T)\;.\label{hcri}\ee  Here $V$ is the volume occupied by the
film,  while the factor $\rho$ is introduced to take into account
the incomplete expulsion of magnetic fields by a thin film, and
the phenomenon of surface nucleation: \be \rho=\frac{\sqrt{24}
\,\lambda}{D}\left(1+\frac{9 D^2}{\pi^6 \xi^2}\right)\;.\ee  For a
film in a cavity, $\Delta F$ is the sum of the condensation energy
${\cal E}_{\rm cond}(T)$ of the film, plus the variation $\Delta
E^{(C)}=E_n^{(C)}-E_s^{(C)}$ of the Casimir (free) energy: \be
\Delta F={\cal E}_{\rm cond}(T)\,+\,\Delta
E^{(C)}(T)\;.\label{varecav}\ee In writing these Equations, we
have exploited the fact that all quantities referring to the film,
like the penetration depth, condensation energy etc. are not
affected by  virtual photons in the surrounding cavity.  This is a
very good approximation, since the leading effect of radiative
corrections is a small  renormalization of the electron mass
\cite{kreuzer} of order $\varepsilon := \alpha \times \hbar
\omega_c/(m c^2)$ (up to logarithmic corrections), where $\alpha$
is the fine structure constant, $\omega_c=c/L$ is the typical
angular frequency of virtual photons, and $m$ is the electron
mass. For $L=10$ nm, $\varepsilon \approx 3 \times 10^{-7}$. The
associated shift of critical field is of the same order of
magnitude, and thus negligible with respect to that caused by
$\Delta E^{(C)}$, which will turn out to be of some percent (see
below).

We see from Eqs. (\ref{hcri})--(\ref{varecav})  that the change in
the Casimir energy causes a shift $\delta H_{c \|}/H_{c \|}
\approx \Delta E^{(C)}/(2{\cal E}_{\rm cond}(T))$ of  critical
field, with respect to its value in a simple film, with same
thickness and temperature. The magnitude of the effect depends on
the relative magnitude of ${\cal E}_{\rm cond}(T)$ and $\Delta
E^{(C)}$, and below we show that these two quantities can indeed
be comparable. Let us begin by the condensation energy ${\cal
E}_{\rm cond}(T)$. As  is well known, it  can be expressed in
terms of the so-called thermodynamical field $H_c(T)$, according
to the formula \cite{tink}   ${\cal E}_{\rm cond}(T)= \;H_c^2(T)
\;V\;/(8 \pi).$ The temperature dependence of the thermodynamical
field approximately follows the parabolic law $H_c(T) \approx
H_c(0)[1-(T/T_c)^2]$. We consider  a film of Beryllium, which is a
type I superconductor with a very low critical temperature
($T_c=24$ mK) and a  low critical field ($H_c(0)=1.08$ Oe)
\cite{soulen}.  Thin Be films possess a much higher critical
temperature, and a proportionally larger thermodynamical field. We
take $D=5$ nm, and then \cite{adams} $T_c \approx 0.5$ K, which
gives $H_c(0) \approx 22.5$ Oe. Thus, using the formula for ${\cal
E}_{\rm cond}(T)$ in terms of $H_c(T)$, and the parabolic law for
$H_c(T)$, we estimate that a Be film, with an area of 1 ${\rm
cm}^2$ (the area is not really important because both the
condensation energy and the Casimir energy are proportional to the
area), at a temperature $T=0.97 \times T_c$ has a condensation
energy ${\cal E}_{\rm cond}(T)\approx 3.5 \times 10^{-8}$ erg. On
the other hand, we see from Eq. (\ref{epara}) that a typical
Casimir energy for a cavity with an area of 1 ${\rm cm}^2$ and a
width $L=10$ nm, has a magnitude of 0.43 erg, i.e.  over ten
million times larger than the condensation energy of the film.  We
see then that a relative variation of Casimir energy as small as
one part in $10^8$  would still correspond to  more than  $10 \%$
of  condensation energy of the film, and would induce a shift of
critical field of over  $5 \%$!

We now have to evaluate    the difference $\Delta E^{(C)}$ among
the Casimir free energies, for the two states of the film.  The
starting point of our analysis is the theory of the Casimir effect
for dispersive media, developed long ago by Lifshitz \cite{lifs}.
In order to establish whether it is applicable to our
superconducting cavity, we briefly recall what are its
assumptions, and what is its range of applicability, as is
obtained from the current literature. The main assumption of the
theory is that, {\it in the relevant range of frequencies and wave
vectors}, one can describe the propagation of electromagnetic
waves inside the media forming the cavity, in terms of a complex
permittivity, depending on the frequency $\omega$ and possibly on
the wave vector $q$.  Thus, provided that one takes into account
the full dependence of the  permittivity on  the wave vector
(besides the frequency), the Lifshitz theory retains its validity
also in cases where space non-local effects become important (see
the discussion in the first of Refs. \cite{bordag}). It is
important to stress that the theory includes also non-retarded
effects \cite{bordag}, and hence it has as limiting cases both van
der Waals forces (that become important at small distances, like
those we consider) and Casimir forces. On this ground, it has been
used recently to study  van der Waals interactions among thin
metal films (of thickness around 10 \AA), till very small
separations (a few \AA) \cite{bostrom}.

It is clear that non-local effects are important, in general, in
superconductors and, for the small separations that we consider
($L=10$ nm), also in normal metals (for an interesting discussion
of non-local effects in the computation of dispersion forces in
superconductors, see Ref. \cite{blos}). However, spatial
dispersion is unimportant for the purpose of computing the {\it
difference} between the Casimir energies in the two states of the
film. The reason is that the optical properties of thin films
(with a thickness $D$ much smaller than the skin depth or
correlation length), in the normal and in the superconducting
states, are indistinguishable for photon energies larger than a
few times $k T_c$ ($k$ being the Boltzmann constant), as accurate
measurements have shown \cite{glover}. This implies that, in the
computation of $\Delta E^{(C)}$, the only relevant photon energies
are those below roughly $10\, k T_c$ (corresponding to the far
IR), which is where the optical properties of the film actually
change when it becomes superconducting. In this region, the
experiments show \cite{glover} that the transmittivity data for
thin superconducting films can be well interpreted in terms of a
complex permittivity that depends only on the frequency, and is
independent of the film thickness.

Starting from the formulae in the first of Ref. \cite{bordag},
that provide a generalization of Lifshitz theory to multilayer
systems, we have obtained the following expression for the
variation of  Casimir energy, in the  limit of low temperatures:
\be \Delta E^{(C)}=\frac{\hbar A}{4 \pi^2 c^2}\int_1^{\infty}
p\,dp \int_0^{\Lambda k T_c/\hbar} d \zeta\,\zeta^2 \,\log
\frac{Q_n^{TE}\,Q_n^{TM}}{Q_s^{TE}\,Q_s^{TM}}\;,\label{lif}\ee
where $\zeta$ is an imaginary freuqency, $p$ is an auxiliary
variable, $\Lambda$ is some cutoff of order 10 or so (the final
results are independent of its precise value) and
\begin{widetext}
\begin{eqnarray}
 Q_I^{TE/TM}( \zeta,p)=
\frac{(1-\Delta_{1I}^{TE/TM}\Delta_{12}^{TE/TM}e^{-2 \zeta p\,
L/c})^2 -(\Delta_{1I}^{TE/TM}-\Delta_{12}^{TE/TM}e^{-2 \zeta \,p\,
L/c})^2 e^{-2 \zeta K_I D/c}}{1-(\Delta_{1I}^{TE/TM})^2 e^{-2
\zeta K_I \,D/c}}\;,\label{delec}\\
\Delta_{j\,l}^{TE}=\frac{K_j- K_l}{K_j+K_l}\;,\;\;
\Delta_{j\,l}^{TM}=\frac{K_j \,\epsilon_l\,(i \zeta)-K_l
\,\epsilon_j\,(i \zeta)}{K_j\, \epsilon_l\,(i \zeta)+K_l\,
\epsilon_j\,(i \zeta)}\;,\;\;K_j=\sqrt{\epsilon_j\,(i
\zeta)-1+p^2}\;,\;\;\;I=n,s\;\;;\;\;j\,,\,l=1,2,n,s.
\end{eqnarray}
\end{widetext}
Here $\epsilon_1=1$, while  $\epsilon_{n/s}(i \zeta)$ denote the
permittivities of the film in the $n/s$ state,  respectively, and
$\epsilon_2(i \zeta)$ is the permittivity of the normal mirrors at
its sides. By using dispersion relations, the dielectric
permittivities $\epsilon_j\,(i \zeta)$ at imaginary frequencies $i
\zeta$ can be expressed in terms of the imaginary part
$\epsilon''(\omega)$ of the dielectric permittivity
$\epsilon(\omega)=\epsilon'(\omega)+i\, \epsilon''(\omega)$ at
real frequencies, i.e. \be \epsilon(i \zeta)-1=\frac{2}{\pi}
\int_0^{\infty} d\omega \frac{\omega
\,\epsilon''(\omega)}{\zeta^2+\omega^2}\;.\label{disp}\ee

For the permittivities $\epsilon_2$  and $\epsilon_n$, we have
used the Drude formula, which is very accurate at low frequencies:
\be \epsilon_j(\omega)=1-\frac{\Omega_{j}^2}{\omega\,(\omega+i/
\tau_j)}\;, \;\;\;\;j=n,2\ee where $\Omega_{j}$ is the
(temperature-independent) plasma frequency, and $\tau_j$ is the
collision time of the metal.  We have neglected the temperature
variation of $\tau_j$, assuming that well before the transition
temperature  they reached their saturation values, as determined
by the impurities present.  We have taken
$\hbar\Omega_{n}=\hbar\Omega_{2}=18.9$ eV (the values for Be) and
$\tau_2=2.4 \times 10^{-12}$ sec (the value for pure bulk Be
samples \cite{soulen}). As for $\tau_n$,  its actual value depends
on the preparation procedure of the film, and as a rule it is much
smaller than in bulk samples. We have considered for it three
values, ranging from $10^{-13}$ sec to $10^{-12}$ sec.
\begin{figure}
\includegraphics{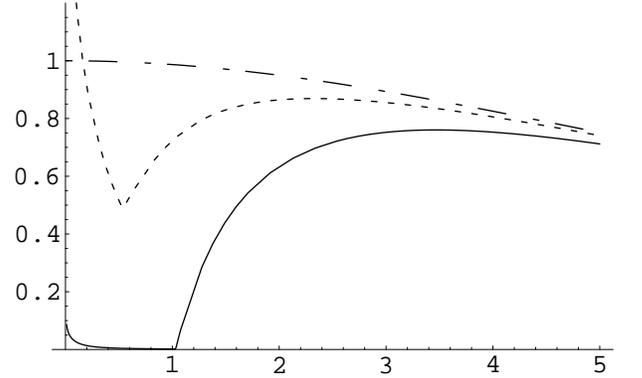}
\caption{\label{fig1} Plots of $  \omega \,\epsilon''_s(\omega)/(
\Omega_n^2 \tau_n)$, for $T/T_c=0.3$ (solid line), $T/T_c=0.9$
(dashed line) and $T=T_c$ (point-dashed line). On the abscissa,
the frequency $\omega$ is in   reduced units $x_0=\hbar \omega/(2
\Delta(0))$ }
\end{figure}

The permittivity $\epsilon_s(i \zeta)$ was computed by
substituting into Eq. (\ref{disp}) the following formula for
$\epsilon''_s(\omega)$ \cite{berl}, which is the long wavelength
limit ($q \rightarrow 0$) of the ordinary Mattis-Bardeen complex
permittivity of   BCS theory: \be \epsilon''_s(\omega)=\frac{\hbar
\Omega^2_n}{2 \omega^2
\tau_n}\left[\int_{\Delta}^{\infty}dE\,J_T+\theta(\hbar \omega-2
\Delta)\int_{\Delta -\hbar \omega}^{-\Delta}dE\,\,J_D
\right]\;,\ee where $\Delta$ is the (temperature dependent) gap
and \begin{eqnarray}
  J_T &:=& \left[\tanh \frac{E+\hbar \omega}{2 k T}-
  \tanh \frac{E}{2kT}\right]\,g(\omega,\tau_n,E) \\
  J_D &:=& -\tanh \left(\frac{E}{2k T}\right)\,g(\omega,\tau_n,E)\;.
\end{eqnarray}
Defining $P_1:=\sqrt{(E+\hbar \omega)^2-\Delta^2}$ and
$P_2:=\sqrt{E^2-\Delta^2}$, the function $g(\omega,\tau_n,E)$ is
\begin{eqnarray}
  g &:=& \left[1+\frac{E(E+\hbar \omega)+\Delta^2}{P_1 P_2}\right]\frac{1}{(P_1-P_2)^2+(\hbar/\tau_n)^2}
  \nonumber\\
   &-&\left[1-\frac{E(E+\hbar \omega)+\Delta^2}{P_1 P_2} \right]\frac{1}{(P_1-P_2)^2+(\hbar/\tau_n)^2}
  \;.\nonumber
\end{eqnarray}
The expression for $\epsilon''(\omega)$ includes also a singular
contribution  at zero frequency $\delta(\omega)/\omega$, with a
coefficient which is determined  so as to satisfy the oscillator
strength sum rule (see Eq. (14) in the first of Refs.
\cite{berl}). Note that for $T \rightarrow T_c$
$\epsilon''_s=\epsilon''_n$  . In Fig. 1, we show the plots of $
\omega \epsilon''_s(\omega)/ ( \Omega_n^2 \tau_n)$, for
$T/T_c=0.3$, $T/T_c=0.9$ and $T=T_c$. The curves are computed for
 $\tau_n=5\times 10^{-13}$ sec. Frequencies are measured in
 reduced units $x_0=\hbar \omega/(2 \Delta(0))$ ($\Delta(0)=7.6 \times 10^{-5}$ eV).
 We have evaluated numerically Eq. (\ref{delec}).  It turns out
 that the contribution of   $TM$ modes is completely negligible
 with respect to that of   $TE$ modes, in agreement with the
 findings of Ref. \cite{bostrom}. For fixed values  of the impurity parameter $y_0=\hbar/(2 \tau_n
\Delta(0))$ (in the range $1< y_0< 30$), $\Delta  E^{(C)}$ has the
following approximate dependence on $L,\, T_c$ and $T/T_c$: \be
\Delta E^{(C)} \propto \frac{1}{L^{0.6}} \times T_c \times
\left(1-\frac{T}{T_c}\right)\;.\label{scale}\ee

\begin{table}
\caption{\label{tab:table1}  Values   of $ \Delta  E^{(C)}$ (in
erg)  for $T/T_c=0.9,\,0.95,\,0.99$, and for   three values of
$\tau_n$ (displayed in the first column). $D=5$ nm, $L=10$ nm,
$A=1\; {\rm cm}^2$.}
\begin{ruledtabular}
\begin{tabular}{cccc}
$\tau_n$ (sec)&    0.9 $T_c$  &  0.95 $T_c$   &  0.99 $T_c$  \\
\hline
$10^{-13}$  & 1.0 $\times 10^{-8}$   &  5.6 $\times 10^{-9}$ & 1.2 $\times 10^{-9}$   \\
$5 \times 10^{-13}$  & 1.9 $\times 10^{-8}$  &1.0 $\times 10^{-8}$  &2.2 $\times 10^{-9}$   \\
$10^{-12}$  &2.15 $\times 10^{-8}$  &  1.2 $\times 10^{-8}$ & 2.5 $\times 10^{-9}$ \\
\end{tabular}
\end{ruledtabular}
\end{table}
\begin{figure}
\includegraphics{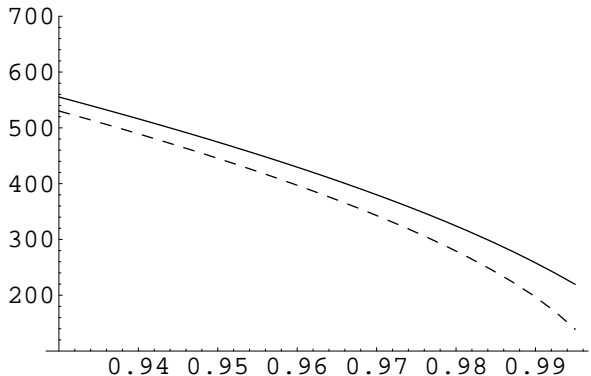}
\caption{\label{fig2} Comparison between  the parallel critical
fields of a Be film in a Casimir cavity (solid curve) and  a
single Be film of same thickness (dashed line),  for $0.93 \le
T/T_c \le 0.995$. The magnetic field is expressed in Oe. $D=5$
{nm}, $L=10$ nm, $\tau_n=5 \times 10^{-13}$ sec.}
\end{figure}
In Table I (last three columns), we report the values of $\Delta
E^{(C)}$, for  three values of $\tau_n$, and for different
temperatures close to $T_c$.  Note that the values of $\Delta
E^{(C)}$ are all $\it  positive$,  and hence the Casimir energy is
smaller in the superconducting state of the film. This implies,
according to Eq. (\ref{hcri}), that the critical magnetic field is
shifted towards larger values. In Fig. 2, we show the parallel
critical field for a Be film placed in a Casimir cavity (solid
line), as compared to that of a simple film (dashed line) of same
thickness, in the interval $0.93 \le T/T_c \le 0.995$, for
$\tau_n=5 \times 10^{-13}$ sec. The curve for the Casimir cavity
has been computed using a power-law fit to the numerical values of
$\Delta E^{(C)}$. As we see from Fig. 2, the shift of  critical
field is larger near $T_c$.

Since close to $T_c$ the thermal fluctuations of the
superconductor order parameter $\psi$ become sensible, one may
wonder whether our results are altered when account is taken of
these fluctuations. We find, however, that the shift of  critical
field resulting from this effect is negligible. The reason is that
{\it the order parameter is confined within the film,  and hence
it is not directly sensitive to the width $L$ of the empty gap at
the sides of the film}. The influence of the cavity width    is
only indirect, and arises from the coupling of electrons in the
film to the virtual photons of the cavity. As we pointed out
earlier (see considerations following Eq. (\ref{varecav}))  this
radiative effect determines a small renormalization  of parameters
in the Ginzburg--Landau free-energy, of order   $ \varepsilon
\approx 3 \times 10^{-7}$ to one-loop accuracy. Since the energy
of thermal fluctuations of $\psi$ in the unperturbed film is at
most of the same order as the condensation energy, we conclude
that the energy shift caused by this effect is of order
$\varepsilon \;{\cal E}_{\rm cond}(T)$, which in turn implies a
shift of critical field $\delta H_c/H_c \approx 10^{-7}$, which is
several orders of magnitude smaller than the shift resulting from
the Casimir energy of virtual photons, that is of some percent.

In conclusion, we find that there is encouraging theoretical
evidence in favor of suitable superconducting cavities being a
promising tool for measuring variations of  Casimir energy. We
think that it would be very interesting to obtain an experimental
verification of the effect of vacuum fluctuations on the critical
field of a Casimir cavity \footnote{Such an experiment, proposed
under the name ALADIN by some of the authors of the present Letter
(G.B., E.C., G.E. and L.R.), jointly with A. Cassinese, F.
Chiarella and R. Vaglio, has been   approved and sponsored by the
INFN.}.

We would like to thank A. Cassinese, F. Tafuri, A. Tagliacozzo and
R. Vaglio for valuable discussions, G.L. Klimchitskaya and V.M.
Mostepanenko for enlightening correspondence. G.B. and G.E.
acknowledge partial financial support by PRIN   {\it SINTESI}.


\begin{thebibliography}{200}



\bibitem{bordag} M. Bordag, U. Mohideen and V.M. Mostepanenko,
Phys. Rep. {\bf 353}, 1 (2001); K. Milton, J. Phys. {\bf A 37},
R209 (2004).




\bibitem{iann} D. Iannuzzi, M. Lisanti and F. Capasso,  Proc. Nat.
Ac. Sci. USA {\bf 101}, 4019 (2004).




\bibitem{glover} R.E. Glover III and M. Tinkham, Phys. Rev. {\bf
108}, 243 (1957).

\bibitem{tink} M. Tinkham, {\it Introduction to Superconductivity}
(McGraw-Hill, Singapore 1996)

\bibitem{kreuzer} M. Kreuzer and K. Svozil, Phys. Rev. {\bf D 34},
1429 (1985).

\bibitem{soulen} R.J. Soulen, Jr., J.H. Colwell and W.E. Fogle, J.
Low Temp. Phys. {\bf 124}, 515 (2001).

\bibitem{adams} P.W. Adams, Phys. Rev. Lett. {\bf 92}, 067003
(2004).



\bibitem{lifs} E.M. Lifshitz, Sov. Phys. JETP {\bf 2}, 73 (1956);
E.M. Lifshitz and L.P. Pitaevskii, {\it Landau and Lifshitz Course
of Theoretical Physics: Statistical Physics Part II}
(Butterworth--Heinemann, 1980.)

\bibitem{bostrom}
M. Bostr$\ddot{\rm o}$m and Bo E. Sernelius,  Phys. Rev. {\bf B
61}, 2204 (2000); ibid. {\bf B 62}, 7523 (2000) and references
therein.



\bibitem{blos} R. Blossey, Europhys. Lett. {\bf 54}, 522 (2001).




\bibitem{berl} A.J.
Berlinsky, C. Kallin, G. Rose and A.C. Shi, Phys. Rev. {\bf B 48},
4074 (1993); W. Zimmermann, E.H. Brandt, M. Bauer, E. Seider and
L. Genzel, Physica C {\bf 183}, 99 (1991).








\end{thebibliography}
\end{document}